\documentclass{jfm}

\usepackage{graphicx}
\usepackage{natbib}
\usepackage{verbatim}

\ifCUPmtlplainloaded \else
  \checkfont{eurm10}
  \iffontfound
    \IfFileExists{upmath.sty}
      {\typeout{^^JFound AMS Euler Roman fonts on the system,
                   using the 'upmath' package.^^J}%
       \usepackage{upmath}}
      {\typeout{^^JFound AMS Euler Roman fonts on the system, but you
                   dont seem to have the}%
       \typeout{'upmath' package installed. JFM.cls can take advantage
                 of these fonts,^^Jif you use 'upmath' package.^^J}%
      }
  \else
  \fi
\fi

\ifCUPmtlplainloaded \else
  \checkfont{msam10}
  \iffontfound
    \IfFileExists{amssymb.sty}
      {\typeout{^^JFound AMS Symbol fonts on the system, using the
                'amssymb' package.^^J}%
       \usepackage{amssymb}%
         
         \let\geq=\geqslant
      }{}
  \fi
\fi

\ifCUPmtlplainloaded \else
  \IfFileExists{amsbsy.sty}
    {\typeout{^^JFound the 'amsbsy' package on the system, using it.^^J}%
     \usepackage{amsbsy}}
    {\providecommand\boldsymbol[1]{\mbox{\boldmath $##1$}}}
\fi

\newsavebox{\astrutbox}
\sbox{\astrutbox}{\rule[-5pt]{0pt}{20pt}}

\title[Various phenomena in fluid mechanics]{Weak compressibility of surface wave
turbulence}

\author[M. Vucelja and I. Fouxon]%
{M\ls A\ls R\ls I\ls J\ls A\ns V\ls U\ls C\ls E\ls L\ls J\ls A$^1$
\and
I\ls T\ls Z\ls H\ls A\ls K\ns F\ls O\ls U\ls X\ls O\ls N$^{1,2}$}

\affiliation{$^1$Physics of Complex Systems, Weizmann Institute
of Science, Rehovot, 76100, Israel\\[\affilskip]
$^2$Racah Institute of Physics, Hebrew University of Jerusalem,
Jerusalem, 91904, Israel}

\begin{document}

\maketitle

\begin{abstract}
We study the growth of small-scale inhomogeneities of the density of
particles floating in weakly nonlinear, small-amplitude, surface
waves. Despite the amplitude smallness, the accumulated effect of
the long-time evolution may produce strongly inhomogeneous
distribution of the floaters: density fluctuations grow
exponentially with a small but finite exponent. We show that the
exponent is of sixth or higher order in wave amplitude.
As a result, the
inhomogeneities do not form within typical time-scales of the
natural environment. We conclude that turbulence of surface waves is
weakly compressible and alone it cannot be a realistic mechanism of
the clustering of matter on liquid surfaces.

\end{abstract}

\section{Introduction}
Clustering of matter on the surface of lakes and pools and of oil
slicks and seaweeds on the sea surface is well-known empirically
while there is no theory that describes it. Since surface flows are
compressible even for incompressible fluids, such theory should be
based on the general description of the development of density
inhomogeneities in a compressible flow. Important characteristics of
the formation of small-scale inhomogeneities is minus the sum of the
Lyapunov exponents of the flow, $\lambda$. It gives the asymptotic
logarithmic growth rate of the density on fluid particle
trajectories at large times. The rate $\lambda$ is minus the average
value of the velocity divergence seen by a fluid particle, and it is
always non-negative because contracting regions with negative
divergence have more particles and hence larger statistical weight,
see \cite{BFF,R1,R2,R}. We note that $\lambda$ is also the
production rate of the Gibbs entropy, so that the condition
$\lambda\geq 0$ can be regarded as an analogue of the second law of
thermodynamics for the dissipative dynamics, see
\cite{R1,R2,R,FF1,FF}. For a generic flow one has $\lambda>0$ and
the asymptotic density becomes a singular measure, the so-called
Sinai-Ruelle-Bowen measure, see e.g. \cite{D}. For floaters this
means that they form a multi-fractal structure on the surface, see
\cite{Yu,FGV,BFF,R,FF1,FF,BGH,BFS,Bruno} for theory and
\cite{gorlum,SO,Sommerer,NAO,alstrom,CG,DFL,Bandi} for experiments.
This structure is the attractor of the two-dimensional dissipative
dynamics obeyed by the particles on the surface, and it evolves
constantly for time-dependent flows, see e.g. \cite{D,O}. The
Kaplan-Yorke dimension of the attractor,
$D_{KY}=1+\lambda_1/|\lambda+\lambda_1|$, is between one and two,
assuming that the principal Lyapunov exponent $\lambda_1$ is
positive, see \cite{O}.

Surface flows are generic compressible flows for which the Eulerian
compressibility, measured by the dimensionless ratio $C=\langle
(\partial_iv_i)^2\rangle/\langle (\partial_jv_i)^2\rangle$, is of
order one, cf. \cite{BDES}. Here angular brackets stand for the
spatial average, $\boldsymbol{v}$ is the floaters velocity field,
and $C$ changes from zero for incompressible flow to one for
potential flow.
One expects $\lambda\sim \lambda_1$ from $C\sim 1$, so that 
the deviation of $D_{KY}$ from the surface dimension $2$ is also of
order one, $2-D_{KY}\sim 1$. The expectation holds true for the flow
on the surface of three-dimensional turbulence. Performing numerical
simulations with the full $3d$ Navier-Stokes equations, \cite{BDES}
found $C\approx 0.5$, $D_{KY}\approx 1.15$ and observed strong
clustering on the surface, see also \cite{CDGS,BDD,Bruno}. However,
underwater turbulence is relatively rare in the natural environment
(due to stable stratification), and it is important to consider
other cases of the surface flows, of which the small-amplitude
surface waves are probably the most wide-spread. Despite the
amplitude smallness, a small-but-finite $\lambda$ produces a large
effect over time-scales of order $1/\lambda$ and larger. Thus to
estimate the role of surface waves in the formation of the floaters
inhomogeneities in the natural environment, one needs to know how
small $\lambda$ is. In this Article we show that $\lambda$ is of
sixth or higher order in wave amplitude. Let us note that for
surface waves the degree of compressibility $C$ is due to linear
waves, which produce potential flow with $C=1$. Thus one could
expect that the estimates $\lambda\sim \lambda_1$ and $2-D_{KY}\sim
1$ would hold for surface waves like they do for the underwater
turbulence. We show that in fact, under some natural non-degeneracy
assumptions described in the Discussion, for surface waves
$\lambda\ll \lambda_1$.

The calculation of $\lambda$ for random waves in different
situations was considered in \cite{BFS,FFV}. For surface waves, in
the linear order in the wave amplitude, the particles move
periodically, see e.g. \cite{Batchelor}, hence there is no net
clustering.
Thus analysis of clustering of the floaters demands the account of
the nonlinear effects.
\cite{BFS} assumed a linear relation between the velocity field of
the floaters and the wave amplitudes, and considered a Gaussian
ensemble of non-interacting waves. The nonlinearity in this case
comes from expressing the Lagrangian objects in terms of the
Eulerian ones. It was shown that $\lambda$ vanishes in the fourth
order in the wave amplitude for longitudinal waves, whose dispersion
relation does not allow the same frequency for two different
wave-vectors (e.g. sound, gravity, capillary waves). On the other
hand, the lowest order non-vanishing contribution into $\lambda_1$
was shown to be always of the fourth order in the wave amplitude.
Under the same assumption of the linear relation between the
velocity and the wave amplitudes, \cite{FFV} demonstrated that the
account of the wave interactions does not change the conclusion of
\cite{BFS} on the vanishing of $\lambda$ in the fourth order in the
wave amplitude.

The results above are inconclusive as far as surface waves are
concerned, for which the relation between the velocity and the
amplitudes is nonlinear due to a small but finite curvature of the
surface, see \cite{ZPhD,Z2}. For this case a separate calculation of
$\lambda$ is needed. Here we provide such a calculation. We consider
weakly nonlinear surface waves and show that neither the wave
interactions, nor the nonlinear relation between the velocity and
the amplitudes, create non-zero sum of the Lyapunov exponents up to
the fourth order in the wave amplitudes. The main tool of our
analysis is a recently derived Green-Kubo type formula for the sum
of the Lyapunov exponents, see \cite{FF,FF1}. This formula expresses
$\lambda$ in terms of the correlations of the flow divergence in the
particle frame. It describes the interplay between the particle
motion and the local flow compressions in accumulating density
inhomogeneities which become pronounced as a result of the long-time
evolution. The slowness of the particle drift from its initial
position allows us to express the correlations in terms of the
Eulerian correlation functions of the velocity, which we further
evaluate by a lengthy, but straightforward calculation, cf.
\cite{FFV}. The following text is organized as follows. In the next
Section we introduce the expression for $\lambda$ in terms of the
Eulerian correlations of the surface flow velocity, valid up to the
fourth order in the wave amplitude. To perform the calculation one
needs to express the velocity in terms of the normal coordinates.
This is dealt with in Section \ref{velocity}. The calculation of the
various terms occurring in $\lambda$ is performed in Section
\ref{Calculation}. Discussion finishes the article.

\section{The sum of the Lyapunov exponents}
\label{Lyapunov}

Behavior of the density $n(t,\boldsymbol{x})$ in a velocity field
$\boldsymbol{v}(t,\boldsymbol{x})$ is governed by the continuity
equation $\partial_t n+\nabla\cdot(\boldsymbol{v} n)=0$, see
\cite{Batchelor}. Introducing Lagrangian particle trajectories by
the equation $\partial _t\boldsymbol{X}(t,\boldsymbol{x})=
\boldsymbol{v}[t,\boldsymbol{X}(t,\boldsymbol{x})]$ with the initial
condition $\boldsymbol{X}(0,\boldsymbol{x})=\boldsymbol{x}$, we may
write the solution for the density as $n[t,
\boldsymbol{X}(t,\boldsymbol{x})]=n(0,
\boldsymbol{x})\exp[-\int_0^{t}w[t',\boldsymbol{X}(t',\boldsymbol{x})]dt']$.
Here we introduced $w(t,\boldsymbol{x})\equiv \boldsymbol{\nabla}
\cdot \boldsymbol{v}(t,\boldsymbol{x})$. We characterize the growth
of spatial inhomogeneities by the asymptotic logarithmic growth rate
$\lambda$ at large times, defined by
\begin{eqnarray}&&
\lambda=\lim_{t\to\infty}\frac{1}{t} \ln \left[\frac{n[t,
\boldsymbol{X}(t,\boldsymbol{x})]}{n(0,
\boldsymbol{x})}\right]=-\lim_{t\to\infty}\frac{1}{t}\int_0^{t}
w[t',\boldsymbol{X}(t',\boldsymbol{x})]dt'.
\end{eqnarray}
The above limit is well-defined, \cite{D}, and it gives minus the
sum of the Lyapunov exponents of the flow
$\boldsymbol{v}(t,\boldsymbol{x})$. It was shown in \cite{FF,FF1}
that if $\boldsymbol{v}(t,\boldsymbol{x})$  is a random, spatially
homogeneous, stationary flow, then one has
\begin{equation}
\lambda=\int^{\infty} _0\! dt\langle w(0,\boldsymbol{x})w
[t,\boldsymbol{X}(t,\boldsymbol{x})] \rangle \,. \label{form1}
\end{equation}
We shall apply the above formula to the case where
$\boldsymbol{v}(t,\boldsymbol{x})$ is the two-dimensional velocity
field governing the motion of the floaters in a (quasi-)stationary
ensemble of weakly nonlinear surface waves sustained by some
forcing, see \cite{ZLF}. We first use the amplitude smallness to
express the Lagrangian correlation function in (\ref{form1}) in
terms of the velocity correlation functions given in the Eulerian
frame. We follow \cite{FFV} who considered (\ref{form1}) in the case
of arbitrary low-amplitude waves. For dispersion relation
$\Omega_{\boldsymbol{k}}$, considering packets with both the
wavenumber and the width of order $k$, the correlation time of $w$
can be estimated as $\Omega_{\boldsymbol{k}}^{-1}$ and the
correlation length as $k^{-1}$. The particle deviation from the
initial position, $\boldsymbol{X}(t,
\boldsymbol{x})-\boldsymbol{x}$, during the period,
$t\simeq\Omega_{\boldsymbol{k}}^{-1}$, is
$\epsilon=kv/\Omega_{\boldsymbol{k}}\ll1$ times smaller than
$k^{-1}$ which allows to expand (\ref{form1}) near $\boldsymbol{x}$.
Performing the expansion to order $\epsilon^4$ we find
\begin{eqnarray}  \label{lambda_init_pos_expansion}
&&\lambda \approx \lambda_2+ \lambda_3+\lambda_4\,,
\\
&& \label{lambda_init_pos_expansion 1} \lambda_2\equiv\int ^\infty_0
dt \langle w(0)w(t)\rangle\,, \quad \lambda_3\equiv\int ^\infty_0
dt\int^{t}_0dt_1 \left\langle w(0) \frac{\partial w(t)}{\partial
x^\alpha} v^\alpha(t_1 )\right\rangle\,,
\\
&& \label{lambda_init_pos_expansion 2}
\lambda_4\equiv\int_0^{\infty}dt\int_0^t\!
dt_1\!\int_0^{t_1}\!dt_2
\left \langle w(0)v_{\beta}(t_2)\left( \frac{\partial w
(t)}{\partial x_{\alpha}}\frac{\partial v_{\alpha} (t_1)}{\partial
x_{\beta}} +\frac{\partial^2 w(t)} {\partial x_{\alpha}\partial
x_{\beta}}v_{\alpha}(t_1)\right)\right\rangle.
\end{eqnarray}
Here and below we suppress the spatial coordinate $\boldsymbol{x}$
over which the averaging is performed. The expansion above was
introduced in \cite{FFV}. We mention that all contributions
$\lambda_i$ are of fourth or higher order in wave amplitude, see
below. To use the above formula to find $\lambda$ to order
$\epsilon^4$, we need to establish the expression for the surface
flow $\boldsymbol{v}$ to order $\epsilon^3$.

\section{The velocity field of the floaters}
\label{velocity}

The velocity field that governs the evolution of the floaters
coordinates $\boldsymbol{r}=(x,y)$ in the horizontal plane, has the
following form
\begin{eqnarray}&&
\boldsymbol{v}(\boldsymbol{r},t)=\left(\frac{\partial
\phi(\boldsymbol{r}, z, t)}{\partial x}\left[z=\eta(\boldsymbol{r},
t)\right], \frac{\partial \phi(\boldsymbol{r}, z, t)}{\partial
y}\left[z=\eta(\boldsymbol{r}, t)\right]\right)\,,
\end{eqnarray}
where $\eta(\boldsymbol{r},t)$ is the surface elevation and
$\phi(\boldsymbol{r}, z, t)$ is the velocity potential,
$\boldsymbol{v}=\nabla \phi$.
\cite{ZPhD, Z2} showed that the system of weakly interacting surface
waves is a Hamiltonian system with canonically conjugate coordinates
$\eta(\boldsymbol{r},t)$ and $\psi(\boldsymbol{r},t)\equiv
\phi(\boldsymbol{r}, \eta(\boldsymbol{r}, t), t)$. Up to the order
$\epsilon^3$, the surface flow of the floaters is given by
\begin{eqnarray}
&&v^\alpha= i\int
\frac{d\boldsymbol{k}_1}{(2\pi)^2}e^{i\boldsymbol{k}_1\cdot
\boldsymbol{r}}k^{\alpha}_1\psi_1 -i\int
\frac{d\boldsymbol{k}_{12}}{(2\pi)^4}
e^{i(\boldsymbol{k}_1+\boldsymbol{k}_2)\cdot \boldsymbol{r}}
|k_1|k^{\alpha}_{2}\psi_1\eta_2
\\ \label{velocity_epsilon_3}
&& -\frac{i}{2}\!\int\!\!\!\frac{d\boldsymbol{k}_{123}}{(2\pi)^6}
e^{i(\boldsymbol{k}_1+\boldsymbol{k}_2+\boldsymbol{k}_3)\cdot
\boldsymbol{r}}
|k_1|\left(|k_1|k^{\alpha}_2+|k_1|k^{\alpha}_3-2\sqrt{k^2_1+k^2_2}k^{\alpha}_3\right)\!\!
\psi_1\eta_2\eta_3\,,
\end{eqnarray}
cf. \cite{Z2} formula (1.8). We introduced shorthand notations
$\eta_i(t)=\eta(\boldsymbol{k}_i,t)$,
$\psi_i(t)=\psi(\boldsymbol{k}_i,t)$ and $d\boldsymbol{k}_{ijl...}=
d\boldsymbol{k}_id\boldsymbol{k}_jd\boldsymbol{k}_l...$. Note that
the velocity field on the surface $\boldsymbol{v}(\boldsymbol{r},t)$
is neither potential, nor solenoidal. We now return to the
expression for the sum of the Lyapunov exponents
(\ref{lambda_init_pos_expansion}).

\section{Calculation of the sum of the Lyapunov exponents}
\label{Calculation}

In this Section we provide the calculation of $\lambda$ to the
fourth order in wave amplitude, based on the calculation of the
different contributions $\lambda_i$, see
(\ref{lambda_init_pos_expansion})-(\ref{lambda_init_pos_expansion
2}). Some parts of this analysis deal with the same objects as those
considered in \cite{BFS,FFV}, however our analysis is different and
it uses specific properties of surface waves. Below we provide a
more detailed calculation, as compared to the short articles
\cite{BFS} and \cite{FFV}.

\subsection{Contribution of the four-point correlation functions}

We first consider the contribution $\lambda_4$ to $\lambda$. To the
fourth order in wave amplitude we may assume Gaussian
non-interacting waves and use Wick's theorem to decouple the
averages. Employing identities like $\langle
v_{\alpha}(t_1)\partial_{\alpha}\partial_{\beta}w(t)\rangle=-
\langle \left(\partial_{\beta}v_{\alpha}(t_1)\right)
\left(\partial_{\alpha}w(t)\right)\rangle$, that follow by
integration by parts, one finds
\begin{eqnarray} \nonumber
&& \lambda_4\!=-\!\int_0^{\infty}\!\!\!dt\!\int_0^t\!\!
dt_1\!\int_0^{t_1}\!\!dt_2\! \Biggl[\!\left\langle w(0)
\frac{\partial w (t)}{\partial x_{\alpha}}\right\rangle\!
\left\langle v_{\alpha} (t_1)w(t_2)\right\rangle +\left\langle
w(0)\frac{\partial v_{\alpha} (t_1)}{\partial
x_{\beta}}\right\rangle \!\!\left\langle w (t)\!\frac{\partial
v_{\beta}(t_2)}{\partial x_{\alpha}}\right\rangle
\\
&& +\left\langle \frac{\partial w(0)} {\partial x_{\alpha}}
\frac{\partial w(t)}{\partial x_{\beta}}\right\rangle \langle
v_{\alpha}(t_1) v_{\beta}(t_2)\rangle +\left\langle w(t_2)
\frac{\partial w (t)}{\partial x_{\alpha}}\right\rangle \left\langle
v_{\alpha}(t_1)w(0)\right\rangle \Biggr]\,. \label{formula1}
\end{eqnarray}
Here we do not assume isotropy of the waves. Isotropy would make
terms like $\langle v_{\alpha} (t_1)w(t_2)\rangle$ vanish. In the
considered order, $\boldsymbol{v}=\nabla \psi$ is a potential field
and spectral representation of the pair-correlation function gives
\begin{eqnarray}\nonumber
&&\langle \psi(0)\psi(t)\rangle=\int
\frac{d\boldsymbol{k}}{(2\pi)^2}E(\boldsymbol{k})\cos\left(\Omega_{\boldsymbol{k}}t\right)
\Rightarrow \langle v_{\alpha}(0)v_{\beta}(t)\rangle=\int
\frac{d\boldsymbol{k}}{(2\pi)^2}k_{\alpha}k_{\beta}E(\boldsymbol{k})
\cos\left(\Omega_{\boldsymbol{k}}t\right),
\end{eqnarray}
and similar expressions for other correlation functions in
(\ref{formula1}). Note that the potentiality of surface waves,
holding in the Gaussian approximation, makes the velocity spectrum
vanish at $k=0$ even if $E(k=0)\neq 0$ [cf. \cite{BFS,FFV}]. We find
\begin{eqnarray} \nonumber
&&\lambda_4=
\int\frac{d\boldsymbol{k}d\boldsymbol{q}E(\boldsymbol{k})E(\boldsymbol{q})}{(2\pi)^4} \int_0^{\infty}dt\int_0^tdt_1\int_0^{t_1}dt_2\Biggl[k^4q^2(\boldsymbol{k}\cdot \boldsymbol{q})
\sin\left[\Omega_{\boldsymbol{k}}t\right]\sin\left[\Omega_{\boldsymbol{q}}(t_1-t_2)\right]
\\
&&-k^2q^2(\boldsymbol{k}\cdot\boldsymbol{q})^2\cos\left[\Omega_{\boldsymbol{k}}t_1\right]
\cos\left[\Omega_{\boldsymbol{q}}(t-t_2)\right]
-k^4(\boldsymbol{k}\cdot\boldsymbol{q})^2\cos\left[\Omega_{\boldsymbol{k}}t\right]
\cos\left[\Omega_{\boldsymbol{q}}(t_1-t_2)\right]\nonumber
\\
&&+k^4q^2(\boldsymbol{k}\cdot{\boldsymbol{q}})\sin\left[\Omega_{\boldsymbol{k}}(t-t_2)\right]
\sin\left[\Omega_{\boldsymbol{q}}t_1\right]\Biggr]. \label{eq221}
\end{eqnarray}
To calculate the time-integrals we represent the products of
trigonometric functions above as sums or differences of cosine
functions and use
\begin{eqnarray}&&
\!\!\int_0^{\infty}\!\!dt\!\int_0^t\! dt_1\!\int_0^{t_1}\!\!dt_2
\cos
(at+bt_1+ct_2)\!=\!-\frac{\pi\delta(a)}{b(b+c)}\!+\!\frac{\pi\delta(a+b)}{bc}
\!-\!\frac{\pi\delta(a+b+c)}{c(b+c)},\ \  b\neq -c, \nonumber\\&&
\int_0^{\infty}\!\!dt\!\int_0^t\! dt_1\!\int_0^{t_1}\!\!dt_2 \cos
(at-bt_1+bt_2)\!=\!-\frac{\pi\delta'(a)}{b}+\frac{\pi\delta(a)}{b^2}-\frac{\pi\delta(a-b)}{b^2}.
\end{eqnarray}
All terms which are supported only at the zero frequency in the
frequency representation ($\delta-$functions or their derivatives)
are also supported only at the zero wavenumber, since the dispersion
relation of surface waves vanishes at $k=0$ only. As a result, due
to the presence of positive powers of $k$ in (\ref{eq221}), these
terms vanish (similarly to the vanishing of the velocity spectrum at
$k=0$ shown above). It is then easy to see that the first and the
fourth terms in $\lambda_4$ (having the same dependence on the
wavevectors) cancel each other, while the second and the third terms
give
\begin{eqnarray}
\label{lambda4_result}
&&
\lambda_4=\!\int\!\!\frac{d\boldsymbol{k}d\boldsymbol{q}}{(2\pi)^4}E(\boldsymbol{k})E(\boldsymbol{q})
k^2(\boldsymbol{k}\cdot\boldsymbol{q})^2(k^2-q^2)\left(\frac{\pi\delta(\Omega_{\boldsymbol{k}}-\Omega_{\boldsymbol{q}})}{2\Omega^2_{\boldsymbol{q}}}\right)\,.
\end{eqnarray}
Since for surface waves the equality of the frequencies of two waves
implies the equality of their wavelengths, then the above terms
cancel each other, $\lambda_4=0$. This reproduces in a simple way
the result of \cite{BFS} which says that in the Gaussian
approximation $\lambda$ vanishes for potential waves having the
property that the equality of the frequencies implies the equality
of the wavelengths.

\subsection{Contribution of the two-point correlation function}

In the next two subsections we calculate those terms in $\lambda_2$
and $\lambda_3$ that can be found directly using Wick's theorem.
After that we calculate the terms involving the wave interactions.
We first consider $\lambda_2$ in (\ref{lambda_init_pos_expansion}).
Using the explicit form of the surface velocity
(\ref{velocity_epsilon_3}) we get
\begin{eqnarray}\nonumber
&&2\lambda_2 = \int
\frac{k_1^2k_2^2d\boldsymbol{k}_{12}dt}{(2\pi)^4}\langle\psi_1(0)\psi_2(t)\rangle
-2\int \frac{d\boldsymbol{k}_{123}dt}{(2\pi)^6}
|k_1|k^2_3(\boldsymbol{k}_1\cdot\boldsymbol{k}_2+k^2_2)
\langle\psi_1(0)\eta_2(0)\psi_3(t) \rangle
\\ \nonumber
&&
-\int \frac{d\boldsymbol{k}_{1234}dt}{(2\pi)^8}
\langle\psi_1(0)\eta_2(0)\eta_3(0)\psi_4(t)\rangle k_4^2|k_1|
\Biggl(|k_1|(\boldsymbol{k}_1\cdot\boldsymbol{k}_2+k^2_2+\boldsymbol{k}_2\cdot\boldsymbol{k}_3)
\\ \nonumber
&&
+\left(|k_1|-2\sqrt{k^2_1+k^2_2}\right)(\boldsymbol{k}_1\cdot\boldsymbol{k}_3+\boldsymbol{k}_2\cdot\boldsymbol{k}_3+k^2_3)\Biggr)+\int
\frac{d\boldsymbol{k}_{1234}dt}{(2\pi)^8}\langle
\psi_1(0)\eta_2(0)\psi_3(t)\eta_4(t)\rangle
\\&& \times
|k_1||k_3|(\boldsymbol{k}_1\cdot\boldsymbol{k}_2+k^2_2)(\boldsymbol{k}_3\cdot\boldsymbol{k}_4+k^2_4)
\end{eqnarray}
We note that the third term above, that can be decomposed by Wick's
theorem, vanishes because it is supported at the zero frequency
$\omega_4$ imposing $k_4=0$. The last term can also be analyzed
using Wick's theorem. Noting that the pair-correlations are
supported at the zero sum of the involved wavenumbers, we find that
the last term equals
\begin{eqnarray}&&
\int \frac{d\boldsymbol{k}_{1234}dt}{(2\pi)^8}\left[\langle
\psi_1(0)\psi_3(t)\rangle\langle\eta_2(0)\eta_4(t)\rangle +\langle
\psi_1(0)\eta_4(t)\rangle\langle\eta_2(0)\psi_3(t)\rangle\right]
\nonumber\\&& \times
|k_1||k_3|(\boldsymbol{k}_1\cdot\boldsymbol{k}_2+k^2_2)(\boldsymbol{k}_3\cdot\boldsymbol{k}_4+k^2_4)\,.
\label{a26}\end{eqnarray} To establish the expressions for the
pair-correlation functions we pass to the normal coordinates
$a(\boldsymbol{k},t)$:
\begin{eqnarray}  \label{psi_a}
\eta(\boldsymbol{k},t) &&=\sqrt{\frac{|k|}{2\Omega_{\boldsymbol{k}}}}\left[a(\boldsymbol{k},t) +a^*(-\boldsymbol{k},t)\right], \,\,
\psi(\boldsymbol{k},t) =-i\sqrt{\frac{\Omega_{\boldsymbol{k}}}{2|k|}} \left[a(\boldsymbol{k},t)-a^*(-\boldsymbol{k},t)\right]\!.
\end{eqnarray}
In the Gaussian approximation the pair correlation functions are
given by
\begin{eqnarray}&&
\langle a^*(\boldsymbol{k}, t)a(\boldsymbol{k}', 0)\rangle
=(2\pi)^2\delta(\boldsymbol{k}-\boldsymbol{k}')n(\boldsymbol{k})
\exp\left[i\Omega_{\boldsymbol{k}}t\right],\ \ \langle
a(\boldsymbol{k}, t)a(\boldsymbol{k}', 0)\rangle=0,
\\&&
\langle \psi(\boldsymbol{k}, t)\psi(\boldsymbol{k}', 0)\rangle=\frac{\Omega_{\boldsymbol{k}}(2\pi)^2\delta(\boldsymbol{k}+\boldsymbol{k}')}{2k}\left[n(\boldsymbol{k})\exp\left(-i\Omega_{\boldsymbol{k}} t\right)+n(-\boldsymbol{k})\exp\left(i\Omega_{-\boldsymbol{k}} t\right)\right],\nonumber\\&&
\langle \eta(\boldsymbol{k}, t)\eta(\boldsymbol{k}',
0)\rangle=\frac{k(2\pi)^2\delta(\boldsymbol{k}+\boldsymbol{k}')}{2\Omega_{\boldsymbol{k}}}\left[n(\boldsymbol{k})\exp\left(-i\Omega_{\boldsymbol{k}} t\right)+n(-\boldsymbol{k})\exp\left(i\Omega_{-\boldsymbol{k}} t\right)\right],\nonumber\\&&
\langle \psi(\boldsymbol{k}, t)\eta(\boldsymbol{k}',
0)\rangle=\frac{(2\pi)^2\delta(\boldsymbol{k}+\boldsymbol{k}')}{2i}\left[n(\boldsymbol{k})\exp\left(-i\Omega_{\boldsymbol{k}} t\right)-n(-\boldsymbol{k})\exp\left(i\Omega_{-\boldsymbol{k}} t\right)\right]\,.
\label{Gaussiancorrelations}
\end{eqnarray}
Using the correlation functions above and noting the vanishing of
the terms containing $\delta-$functions supported at the zero
frequency, we can write (\ref{a26}) as
\begin{eqnarray}&&
\int \frac{d\boldsymbol{k}_{12}}{(2\pi)^{3}}\Biggl[k_1^2(\boldsymbol{k}_1\cdot\boldsymbol{k}_2+k^2_2)^2
\left(\frac{\Omega_1k_2}{4k_1\Omega_2}\right)\left[n_1n_{-2}\delta(\Omega_1-\Omega_{-2})
+n_{-1}n_2\delta(\Omega_{-1}-\Omega_2)\right]\nonumber
\\&&-\frac{1}{4}
|k_1||k_2|(\boldsymbol{k}_1\cdot\boldsymbol{k}_2+k^2_2)(\boldsymbol{k}_1\cdot\boldsymbol{k}_2+k^2_1)\left[n_1n_{-2}\delta(\Omega_1-\Omega_{-2})+n_{-1}n_2
\delta(\Omega_{-1}-\Omega_2)\right]\Biggr]=0,
\nonumber\end{eqnarray} where we introduced the shorthand notation:
$n(\pm\boldsymbol{k}_i,t)=n_{\pm i}(t)$ and $\Omega_{\pm
\boldsymbol{k}_i}=\Omega_{\pm i}$. We use that for surface waves
$\Omega_{\boldsymbol{k}}$ is an increasing function of
$|\boldsymbol{k}|$. We also use that $\delta-$functions imply
$\Omega_1=\Omega_2$ and $k_1^2=k_2^2$. We find that $\lambda_2$ can
be written as
\begin{equation} \label{lambda2_int}
\lambda_2=\!\!\int\!\!\frac{d\boldsymbol{
k}_{12}dt}{(2\pi)^4}\left[\frac{k_1^2k_2^2}{2}\langle
\psi_1(0)\psi_2(t)\rangle-\!\int\!\!
\frac{d\boldsymbol{k}_{3}|k_1|k^2_3(\boldsymbol{k}_1\!\cdot\!\boldsymbol{k}_2+k^2_2)}{(2\pi)^{d}}\!
\langle \psi_1(0)\eta_2(0)\psi_3(t) \rangle\right]\!.
\end{equation}
The calculation of the above terms demands the account of the
interactions. We now consider the terms in $\lambda_3$ that can be
calculated by the direct use of Wick's theorem.

\subsection{Contribution of the three-point correlation function}

We consider $\lambda_3$ from (\ref{lambda_init_pos_expansion}).
Using the expression (\ref{velocity_epsilon_3}) for the velocity, we
obtain
\begin{eqnarray}&& \label{lambda3_int}
\lambda_3=-\int \frac{d\boldsymbol{k}_{123}}{(2\pi)^6}k^2_1k^2_2(\boldsymbol{k}_2\cdot\boldsymbol{k}_3)\int^\infty _0 dt \int ^t _0 dt_1 \langle\psi_1(0)\psi_2(t)\psi_3(t_1) \rangle \nonumber\\&&
+\int \frac{d\boldsymbol{k}_{1234}}{(2\pi)^8}\int^\infty _0 dt \int ^t _0 dt_1\bigg[ |k_1|(\boldsymbol{k}_1\cdot\boldsymbol{k}_2+k^2_2)k^2_3(\boldsymbol{k} _3\cdot\boldsymbol{k} _4)
\langle\psi_1(0)\eta_2(0)\psi_3(t)\psi_4(t_1)\rangle
\nonumber
\\&&
+k^2_1|k_2|(\boldsymbol{k}_2\cdot\boldsymbol{k}_3+k^2_3)(\boldsymbol{k}_2\cdot\boldsymbol{k}_4+\boldsymbol{k}_3\cdot\boldsymbol{k}_4)
\langle\psi_1(0)\psi_2(t)\eta_3(t)\psi_4(t_1) \rangle \nonumber
\\&&
+k^2_1 k^2_2(\boldsymbol{k}_2\cdot\boldsymbol{k}_4)|k_3|
\langle \psi_1(0) \psi_2(t)\psi_3(t_1)\eta_4(t_1)\rangle \bigg].
\label{lambda3}
\end{eqnarray}
One can use Wick's theorem for the last three terms. Employing the
identity
\begin{eqnarray}&&
\int ^\infty _0 dt \int ^t _0 dt_1 \exp[i\omega_1t+i\omega_2t_1]=
\frac{i\pi[\delta(\omega_1)-\delta(\omega_1+\omega_2)]}{\omega_2},
\label{identity}\end{eqnarray} one finds that part of the obtained
terms contain $\delta-$functions supported at the zero frequency and
they vanish because of the vanishing of the integrand there. Let us
consider the rest of the terms. For the first of the fourth-order
terms in (\ref{lambda3}) one finds
\begin{eqnarray}
\nonumber
&&
\int\frac{d\boldsymbol{k}_{12}}{(2\pi)^4} \frac{\Omega_1\pi}{4k_1}(\boldsymbol{k} _1\cdot\boldsymbol{k}
_2)
\Biggl\lbrace
|k_1|^3(\boldsymbol{k}_1\cdot\boldsymbol{k}_2+k^2_2)
\Biggl[ \frac{ n_{-1}n_2\delta(\Omega_{-1}-\Omega_2)}{\Omega_2}
+\frac{n_{-2}n_1\delta(\Omega_{-2}-\Omega_1)}{\Omega_{-2}}\Biggr]
\\
\nonumber
&&
-|k_1| k_2^2(\boldsymbol{k}_1\cdot\boldsymbol{k}_2+k^2_2)
\Biggl[\frac{n_{-1}n_2\delta(\Omega_{-1}-\Omega_2)}{\Omega_{-1}}
+\frac{n_{-2}n_1\delta(\Omega_{-2}-\Omega_1)}{\Omega_{1}} \Biggr]\Biggr\rbrace=0,
\end{eqnarray}
where the equality of the frequencies implies the equality of the
wavelengths. Analogously, for the second Gaussian term one finds
\begin{eqnarray}
\nonumber
&&
\int \frac{d\boldsymbol{k}_{23}}{(2\pi)^4} \frac{\Omega_2\pi}{4k_2}
\Biggl\lbrace
|k_2|^3(\boldsymbol{k}_2\cdot\boldsymbol{k}_3+k^2_3)^2
\Biggl[
 \frac{ n_{-3}n_2\delta(\Omega_{-3}-\Omega_2)}{\Omega_{-3}}
+\frac{n_{-2}n_3\delta(\Omega_{-2}-\Omega_3)}{\Omega_{3}} \Biggr]
\\ \nonumber
&&
-|k_2| k_3^2(\boldsymbol{k}_2\cdot\boldsymbol{k}_3+k^2_3)(\boldsymbol{k}_2\cdot\boldsymbol{k}_3+k^2_2)\Biggl[\frac{
n_{-3}n_2\delta(\Omega_{-3}-\Omega_2)}{\Omega_{2}}
+\frac{n_{-2}n_3\delta(\Omega_{-2}-\Omega_3)}{\Omega_{-2}}
\Biggr]\Biggr\rbrace\!=0.
\end{eqnarray}
Finally, the third Gaussian term contains only $\delta-$functions
supported at the zero frequencies, so it also produces zero.

\subsection{Contribution of the interaction terms}

As a result of the calculation in the previous sections, adding
equations (\ref{lambda2_int}) and (\ref{lambda3_int}), one can write
$\lambda$ as a sum over terms which calculation involves the wave
interactions.
 These terms have a special property important for the following
analysis: all the terms contain a field at the zero frequency. Using
Fourier representation over the frequency one finds
\begin{eqnarray} \nonumber
&& \lambda=\frac{1}{2}\int
\frac{k_1^2k_2^2d\boldsymbol{k}_{12}d\omega}{(2\pi)^5}\langle
\psi_1(\omega)\psi_2(\omega=0)\rangle
\\
\nonumber
&&
-\int \frac{d\boldsymbol{k}_{123}d\omega_{12}}{(2\pi)^8} |k_1|k^2_3(\boldsymbol{k}_1\cdot\boldsymbol{k}_2+k^2_2)
\langle \psi_1(\omega_1)\eta_2(\omega_2)\psi_3(\omega=0) \rangle
\\
\label{lambdaexpression}
&&
-\int \frac{d\boldsymbol{k}_{123}d\omega_{123}}{(2\pi)^9}k^2_1k^2_2(\boldsymbol{k}_2\cdot\boldsymbol{k}_3) \langle \psi_1(\omega_1)\psi_2(\omega_2)\psi_3(\omega_3) \rangle
\frac{i\pi[\delta(\omega_2)-\delta(\omega_1)]}{\omega_3},
\end{eqnarray} where we introduced shorthand notation
$d\omega_{ijl}=d\omega_id\omega_jd\omega_l...$ and used in the last
term the identity (\ref{identity}) and the proportionality of the
correlation function to $\delta(\omega_1+\omega_2+\omega_3)$. We
observe that all the terms contain $\psi(\boldsymbol{k}, \omega=0)$.
We assume that the forcing sustaining the stationary wave turbulence
vanishes at the zero frequency [note that the first term on the RHS
above vanishes in the Gaussian approximation independently of this
assumption, cf. \cite{BFS} and \cite{FFV}] and express
$\psi(\boldsymbol{k}, \omega=0)$ with the help of higher order
terms. Substituting the resulting expressions into the correlation
functions will allow the use of Wick's decomposition.

\subsubsection{The expression for $\psi(\boldsymbol{k}, \omega=0)$}

To calculate $\lambda$ we need to know $\psi(\boldsymbol{k},
\omega=0)$ to the third order in wave amplitude. Consider the
dynamical equation on the surface elevation $\eta$, see
\cite{Batchelor,Z2},
\begin{eqnarray}&&
\frac{\partial \eta}{\partial t}=\frac{\partial \phi}{\partial
z}[z=\eta]-\nabla \eta\nabla\phi [z=\eta].
\label{dynamics0}\end{eqnarray} To order $\epsilon^2$, the equation
reads
\begin{eqnarray}&&
\frac{\partial \eta}{\partial t}=\frac{\partial \phi_0}{\partial
z}[z=0]+\eta\frac{\partial^2\phi_0}{\partial
z^2}[z=0]+\frac{\partial \phi_1}{\partial z}-\nabla
\eta\nabla\phi_0[z=0]+O(\eta^3), \label{dynamics}\end{eqnarray}
where $\phi_0$ and $\phi_1$ are the terms of the expansion of the
potential with respect to the surface elevation, see \cite{Z2}.
Using the expressions for $\phi_i$, see \cite{Z2}, performing the
Fourier transform over the space and time coordinates, and setting
the frequency $\omega=0$, we find
\begin{eqnarray}&&
0=|k|\psi(\boldsymbol{k}, \omega=0)
+\int\frac{ d\boldsymbol{k}_1d\omega}{(2\pi)^3}k_1^2\psi(\boldsymbol{k}_1, \omega)
\eta(\boldsymbol{k}-\boldsymbol{k}_1, -\omega)
-\int\frac{ d\boldsymbol{k}_1d\omega}{(2\pi)^3}|k||k_1|\psi(\boldsymbol{k}_1, \omega)\nonumber
\\&& \times\eta(\boldsymbol{k}-\boldsymbol{k}_1, -\omega)
+\int\frac{ d\boldsymbol{k}_1d\omega}{(2\pi)^3}\boldsymbol{k}_1\cdot(\boldsymbol{k}-\boldsymbol{k}_1)\psi(\boldsymbol{k}_1, \omega)\eta(\boldsymbol{k}-\boldsymbol{k}_1, -\omega)+O(\eta^3).
\end{eqnarray}
It follows that
\begin{eqnarray}&&
\psi(\boldsymbol{k}, \omega=0)=\int\frac{
d\boldsymbol{k}_1d\omega}{(2\pi)^3}\left(|k_1|-\frac{\boldsymbol{k}_1\cdot\boldsymbol{k}}{|k|}\right)
\psi(\boldsymbol{k}_1, \omega)\eta(\boldsymbol{k}-\boldsymbol{k}_1,
-\omega)+O(\eta^3). \label{zerofreq}\end{eqnarray} The physical
meaning of the above representation is that the zero frequency field
arises due to the nonlinear interactions only, and in the lowest
order it can be represented as a result of a single wave scattering.
The above formula is sufficient to calculate the interaction terms
containing the products of three fields.

\subsubsection{Interaction terms containing the products of three fields}

To calculate the interaction terms of the third order we use
(\ref{zerofreq}), Wick's decomposition and Fourier transformed
version of (\ref{Gaussiancorrelations}). The second term in
(\ref{lambdaexpression}) can be written with the help of
(\ref{zerofreq}) as
\begin{eqnarray}&&
-\int \frac{d\boldsymbol{k}_{1234}d\omega_{123}}{(2\pi)^{11}}
|k_1|k^2_3(\boldsymbol{k}_1\cdot\boldsymbol{k}_2+k^2_2)\left(|k_4|-\frac{\boldsymbol{k}_4\cdot\boldsymbol{k}_3}{|k_3|}\right)\langle \psi(\boldsymbol{k}_1,
\omega_1)\eta(\boldsymbol{k}_2, \omega_2) \psi(\boldsymbol{k}_4,
\omega_3)\nonumber\\&& \eta(\boldsymbol{k}_3-\boldsymbol{k}_4, -\omega_3)
\rangle=-\int \frac{d\boldsymbol{k}_{12}d\omega_{1}}{(2\pi)^{7}}|k_1|(\boldsymbol{k}_1+\boldsymbol{k}_2)^2(\boldsymbol{k}_1\cdot\boldsymbol{k}_2+k^2_2)\left(|k_1|-\frac{\boldsymbol{k}_1\cdot(\boldsymbol{k}_1+\boldsymbol{k}_2)}{|\boldsymbol{k}_1+\boldsymbol{k}_2|}\right)\nonumber
\\&&
\left(\frac{\Omega_1|k_2|}{4\Omega_2|k_1|}\right)\left[n_1\delta(\omega_1+\Omega_1)+n_{-1}
\delta(\omega_1-\Omega_{-1})\right]\left[n_2\delta(-\omega_1+\Omega_2)+n_{-2}
\delta(\omega_1+\Omega_{-2})\right]\nonumber
\\&&
+\frac{1}{4}\int\frac{d\boldsymbol{k}_{12}d\omega_{1}}{(2\pi)^{7}}|k_1|
(\boldsymbol{k}_1+\boldsymbol{k}_2)^2(\boldsymbol{k}_1\cdot\boldsymbol{k}_2+k^2_2)
\left(|k_2|-\frac{\boldsymbol{
k}_2\cdot(\boldsymbol{k}_1+\boldsymbol{k}_2)}{|\boldsymbol{k}_1+\boldsymbol{k}_2|}\right)\nonumber\\&&\times
\left[n_1\delta(\omega_1+\Omega_1)-n_{-1}
\delta(\omega_1-\Omega_{-1})\right]\left[n_{-2}\delta(\omega_{1}+\Omega_{-2})-n_{2}
\delta(\omega_{1}-\Omega_{2})\right],\end{eqnarray} where to show
that the remaining contraction vanishes, one should use that
$\langle \psi(\boldsymbol{k}, \omega=0)\rangle$ is representable as
an integral over $\langle \psi(\boldsymbol{k},
t)\eta(\boldsymbol{k}', t)\rangle$ which in the Gaussian
approximation vanishes by (\ref{Gaussiancorrelations}). Using
$\Omega_{-\boldsymbol{k}}=\Omega_{\boldsymbol{k}}$ and noting that
the terms supported at $\Omega_{\boldsymbol{k}}=0$ vanish, one may
rewrite the above as
\begin{eqnarray}&&
-\frac{1}{4}\int \frac{d\boldsymbol{k}_{12}}{(2\pi)^{7}}|k_1|(\boldsymbol{k}_1+\boldsymbol{k}_2)^2
(\boldsymbol{k}_1\cdot\boldsymbol{k}_2+k^2_2)\left(|k_1|
-\frac{\boldsymbol{k}_1\cdot(\boldsymbol{k}_1+\boldsymbol{k}_2)}
{|\boldsymbol{k}_1+\boldsymbol{k}_2|}\right)
\Biggl[n_1n_{-2}\delta(\Omega_1-\Omega_{-2})\nonumber
\\
&&
+n_{-1}n_2\delta(\Omega_2-\Omega_{-1})\Biggr]+\frac{1}{4}\int \frac{d\boldsymbol{k}_{12}}{(2\pi)^{7}}|k_1|(\boldsymbol{k}_1+\boldsymbol{k}_2)^2
(\boldsymbol{k}_1\cdot\boldsymbol{k}_2+k^2_2)\left(|k_2|
-\frac{\boldsymbol{k}_2\cdot(\boldsymbol{k}_1+\boldsymbol{k}_2)}{|\boldsymbol{k}_1
+\boldsymbol{k}_2|}\right)\nonumber
\\
&&\left[n_1n_{-2}\delta(\Omega_1-\Omega_{-2})+n_{-1}n_2
\delta(\Omega_2-\Omega_{-1})\right]=\frac{1}{4}\int \frac{d\boldsymbol{k}_{12}}{(2\pi)^{7}}|k_1|(\boldsymbol{k}_1+\boldsymbol{k}_2)^2(\boldsymbol{k}_1\cdot
\boldsymbol{k}_2+k^2_2)
\nonumber
\\
&&
\left(\frac{(\boldsymbol{k}_1-\boldsymbol{k}_2)\cdot(\boldsymbol{k}_1+\boldsymbol{k}_2)}{|\boldsymbol{k}_1+\boldsymbol{k}_2|}\right)
\Biggl[n_1n_{-2}\delta(\Omega_1-\Omega_{-2})+n_{-1}n_2
\delta(\Omega_2-\Omega_{-1})\Biggr]=0 ,\end{eqnarray} where the last
term contains $k_1^2-k_2^2=0$. Let us now consider the last term in
(\ref{lambdaexpression}) that can be written as
\begin{eqnarray}&&
i\pi\int \frac{d\boldsymbol{k}_{123}d\omega_{23}}{\omega_3(2\pi)^{9}}k^2_1k^2_2\boldsymbol{k}_3\cdot(\boldsymbol{k}_2-\boldsymbol{k}_1) \langle \psi(\boldsymbol{k}_1,
\omega=0)\psi(\boldsymbol{k}_2, \omega_2)\psi(\boldsymbol{k}_3, \omega_3)
\rangle.\end{eqnarray} Substituting (\ref{zerofreq}) we find
\begin{eqnarray}&&
i\pi\!\int\! \frac{d\boldsymbol{k}_{1234}d\omega_{234}k^2_1k^2_2\boldsymbol{k}_3\!\cdot\!(\boldsymbol{k}_2-\boldsymbol{k}_1)}
{\omega_3(2\pi)^{12}}\! \left(|k_4|
-\frac{\boldsymbol{k}_4\cdot\boldsymbol{k}_1}{|k_1|}\right)\!
\langle \eta(\boldsymbol{k}_1-\boldsymbol{k}_4,-\omega_4)\psi(\boldsymbol{k}_2, \omega_2)
\psi(\boldsymbol{k}_3,\omega_3)\nonumber\\
&&\times\psi(\boldsymbol{k}_4, \omega_4)
\rangle=i\pi\!\int\! \frac{d\boldsymbol{k}_{23}d\omega_{3}}{\omega_3(2\pi)^{8}}(\boldsymbol{k}_2
+\boldsymbol{k}_3)^2k^2_2\boldsymbol{k}_3\cdot(2\boldsymbol{k}_2+\boldsymbol{k}_3)
\left(|k_3|-\frac{\boldsymbol{k}_3\cdot(\boldsymbol{k}_2+\boldsymbol{k}_3)}{|\boldsymbol{k}_2
+\boldsymbol{k}_3|}\right)\frac{\Omega_3}{4i|k_3|}\nonumber\\
&&\left[n_2\delta(\omega_3-\Omega_2)-n_{-2}\delta(\omega_3+\Omega_{-2})\right]
\left[n_{3}\delta(\omega_{3}+\Omega_{3})+n_{-3}
\delta(\omega_{3}-\Omega_{-3})\right]+i\pi\int \frac{d\boldsymbol{k}_{23}d\omega_{3}}{\omega_3(2\pi)^{8}}\nonumber\\
&& \times(\boldsymbol{k}_2+\boldsymbol{k}_3)^2k^2_2\boldsymbol{k}_3\cdot(2\boldsymbol{k}_2+\boldsymbol{k}_3)
\left(|k_2|-\frac{\boldsymbol{k}_2\cdot(\boldsymbol{k}_2+\boldsymbol{k}_3)}{|\boldsymbol{k}_2+
\boldsymbol{k}_3|}\right)\left(\frac{\Omega_2}{4i|k_2|}\right)\nonumber\\
&&\times\left[n_{2}\delta(\omega_{3}-\Omega_{2})+n_{-2}
\delta(\omega_{3}+\Omega_{-2})\right]\left[n_3\delta(\omega_3+\Omega_3)-n_{-3}
\delta(\omega_3-\Omega_{-3})\right],
\end{eqnarray}
where the term involving the contraction $\langle
\eta(\boldsymbol{k}_1-\boldsymbol{k}_4,-\omega_4)\psi(\boldsymbol{k}_4,
\omega_4) \rangle$ corresponding to $\langle \psi(\boldsymbol{k},
\omega=0)\rangle$, vanishes as was shown in the analysis of the
previous term, where we omitted the terms supported at the zero
frequency. Taking the integral over $\omega_3$, omitting the terms
supported at the zero frequency, we find
\begin{eqnarray}&&
i\pi\int \frac{d\boldsymbol{k}_{23}}{(2\pi)^{8}}(\boldsymbol{k}_2+\boldsymbol{k}_3)^2k^2_2\boldsymbol{k}_3\cdot(2\boldsymbol{k}_2+\boldsymbol{k}_3)
\left(|k_3|-\frac{\boldsymbol{k}_3\cdot(\boldsymbol{k}_2+\boldsymbol{k}_3)}{|\boldsymbol{k}_2+\boldsymbol{k}_3|}\right)\nonumber\\&&\left[\frac{n_2n_{-3}
\delta(\Omega_{2}-\Omega_{-3})}{4i|k_3|}+\frac{n_{-2}n_{3}\delta(\Omega_{-2}-\Omega_{3})}{4i|k_3|}\right]+i\pi\int
\frac{d\boldsymbol{k}_{23}}{(2\pi)^{8}}\nonumber\\&& \times(\boldsymbol{k}_2+\boldsymbol{k}_3)^2k^2_2\boldsymbol{k}_3\cdot(2\boldsymbol{k}_2+\boldsymbol{k}_3)
\left(|k_2|-\frac{\boldsymbol{k}_2\cdot(\boldsymbol{k}_2+\boldsymbol{k}_3)}{|\boldsymbol{k}_2+\boldsymbol{k}_3|}\right)\nonumber\\&&\left[-\frac{n_{-2}n_3\delta(\Omega_{3}-\Omega_{-2})}
{4i|k_2|}-\frac{n_{2}n_{-3}\delta(\Omega_{2}-\Omega_{-3})}{4i|k_2|}\right]=0,
\end{eqnarray}
where zero is obtained in the same way as with the previous term.
The result of this subsection is that $\lambda$ can be written as
\begin{eqnarray}&&
\lambda=\frac{1}{2}\int \frac{k_1^2k_2^2d\boldsymbol{k}_{12}}{(2\pi)^4}\int\frac{d\omega}{2\pi}\langle \psi(\boldsymbol{k}_1,
\omega)\psi(\boldsymbol{k}_2, \omega=0)\rangle.
\label{lambdalast}\end{eqnarray}

\subsubsection{Interaction term containing the product of two fields}

To calculate the RHS of (\ref{lambdalast}) we note that it is
sufficient to know $\psi(\boldsymbol{k}_1, \omega)$ at an
arbitrarily small but finite $\omega$ where the forcing is again
negligible and one use the dynamic equation (\ref{dynamics0})
without the force. The Fourier transform of (\ref{dynamics}) taken
now at a small but finite frequency gives
\begin{eqnarray}
\nonumber
&&
i\omega\eta(\boldsymbol{k}, \omega)=|k|\psi(\boldsymbol{k}, \omega)+\int\frac{
d\boldsymbol{k}_1d\omega_1}{(2\pi)^3}k_1^2\psi_1(\omega_1)\eta(\boldsymbol{k}-\boldsymbol{k}_1, \omega-\omega_1)-\int\!\frac{ d\boldsymbol{k}_1d\omega_1}{(2\pi)^3}|k||k_1|\psi_1(\omega)
\\&&
\times \eta(\boldsymbol{k}-\boldsymbol{k}_1, \omega-\omega_1)
+\int\frac{ d\boldsymbol{k}_1d\omega_1}{(2\pi)^3}\boldsymbol{k}_1\cdot(\boldsymbol{k}-\boldsymbol{k}_1)
\psi_1(\omega)\eta(\boldsymbol{k}-\boldsymbol{k}_1, \omega-\omega_1)+O(\eta^3).
\end{eqnarray}
which produces
\begin{eqnarray}&&
\psi(\boldsymbol{k}, \omega)=-\frac{i\omega\eta(\boldsymbol{k},
\omega)}{|k|}+\int\frac{ d\boldsymbol{k}_1d\omega_1}{(2\pi)^3}\left(|k_1|-\frac{\boldsymbol{k}_1\cdot\boldsymbol{k}}{|k|}\right)\psi(\boldsymbol{k}_1, \omega_1)\eta(\boldsymbol{k}-\boldsymbol{k}_1,
\omega-\omega_1)+O(\eta^3). \nonumber\end{eqnarray}
It follows that the (\ref{lambdalast}) can be written as
\begin{eqnarray}&&
\lambda=\frac{1}{2}\int \frac{k_1^2k_2^2d\boldsymbol{k}_{12}}{(2\pi)^4}\int\frac{d\omega}{2\pi}\Biggl\langle
\Biggl[-\frac{i\omega\eta(\boldsymbol{k}_1, \omega)}{|k_1|}+\int\frac{
d\boldsymbol{k}_3d\omega_1}{(2\pi)^3}\left(|k_3|-\frac{\boldsymbol{k}_3\cdot\boldsymbol{k}_1}{|k_1|}\right)\psi(\boldsymbol{k}_3, \omega_1)\nonumber\\&&
\times\eta(\boldsymbol{k}_1-\boldsymbol{k}_3,
\omega-\omega_1)+O(\eta^3)\Biggr]\psi(\boldsymbol{k}_2,
\omega=0)\Biggr\rangle=\frac{1}{2}\int \frac{k_1^2k_2^2d\boldsymbol{k}_{12}}{(2\pi)^4}\int\frac{d\omega_1}{2\pi}\Biggl\langle
\Biggl[\int\frac{ d\boldsymbol{k}_3d\omega_3}{(2\pi)^3}\nonumber\\&&
\times\left(|k_3|-\frac{\boldsymbol{k}_3\cdot\boldsymbol{k}_1}{|k_1|}\right)\psi(\boldsymbol{k}_3, \omega_3)
\eta(\boldsymbol{k}_1-\boldsymbol{k}_3, \omega_1-\omega_3)+O(\eta^3)\Biggr]\psi(\boldsymbol{k}_2,
\omega=0)\Biggr\rangle,
\end{eqnarray}
where we used that $ \omega\langle\eta(\boldsymbol{k}_1,
\omega)\psi(\boldsymbol{k}_2, \omega=0)\rangle\propto
\omega\delta(\omega)=0$. Finally, using that $\psi(\boldsymbol{k},
\omega=0)$ is by itself of order $\epsilon^2$, see (\ref{zerofreq}),
taking the integral over $\omega_3$ and omitting the terms supported
at the zero frequency, we find $\lambda$ to the order $\epsilon^4$
\begin{eqnarray}
\nonumber
&&
\lambda=
\frac{1}{8}\int \frac{d\boldsymbol{k}_{13}}{(2\pi)^{3}} (\boldsymbol{k}_1+\boldsymbol{k}_3)^4
\left(|k_3|-\frac{\boldsymbol{k}_3\cdot(\boldsymbol{k}_1+\boldsymbol{k}_3)}
{|\boldsymbol{k}_1+\boldsymbol{k}_3|}\right)
[n_3n_{-1}\delta(\Omega_{-1}-\Omega_3)
\\
&& +n_{-3}n_1\delta(\Omega_1-\Omega_{-3})]
\left(|k_3|-|k_1|-\frac{(\boldsymbol{k}_3-\boldsymbol{k}_1)
\cdot(\boldsymbol{k}_1+\boldsymbol{k}_3)}{|\boldsymbol{k}_1+\boldsymbol{k}_3|}\right)=0.
\end{eqnarray}
The above can be easily verified by noting that the integrand can be
reduced to an expression proportional to $k_1^2-k_3^2$. Thus we
obtain that the sum of Lyapunov exponents for weakly interacting
surface waves is zero in the fourth order in the wave amplitude.

\section{Discussion}
\label{discussion}

We have shown that the sum of the Lyapunov exponents for surface
wave turbulence vanishes in the fourth order in wave amplitude.
Using the approximate Gaussianity of the statistics, it is easy to
see that the leading order term in $\lambda$ is of the sixth order
in wave-amplitude, or higher. Therefore we have derived that
$\lambda\lesssim \Omega _{\boldsymbol{k}}\epsilon^6$. For waves with
a typical period of the order of seconds and with not too small
$\epsilon=0.1$, we find that the time-scale $1/\Omega
_{\boldsymbol{k}}\epsilon^6$  is of the order of weeks. Thus, even
if there is no degeneracy in the sixth order and $\lambda\sim \Omega
_{\boldsymbol{k}}\epsilon^6$, the formation of the inhomogeneities
would occur at the time-scale of weeks and larger. It is unlikely
that a low-amplitude wave turbulence would persist for such time.
Thus we expect the turbulence of small-amplitude surface waves to
have negligible effect on the formation of the floaters
inhomogeneities in realistic situations. Let us stress that this
weak compressibility of surface waves holds in the sense of the
long-time action of the flow on the particles, while the
characteristic value of the ratio of the surface flow divergence to
the curl is of order one.

Let us consider some estimates for the Lyapunov exponents of the
floaters velocity field. For the non-interacting Gaussian surface
waves, $\lambda_1$ is non-zero in the fourth order in wave
amplitude, see \cite{BFS}, while $\lambda$ is non-zero in the sixth
order, \cite{BFS,MGS}. The nonlinear wave interactions and the
nonlinearity of the velocity-amplitude relation add to $\lambda_1$
additional terms of the fourth order in the wave amplitude and
higher. It is highly implausible that these terms produce exact
cancelation of $\lambda_1$ in the fourth order - such cancelation
would depend on the precise form of both the interactions and the
velocity-amplitude relation. Moreover, positive Lyapunov exponent
and Lagrangian chaos hold even for rather simple Eulerian flows, see
e.g. \cite{BH}, so no degeneracy for $\lambda_1$ is expected that
would lead to an exact cancelation in the fourth order. Therefore we
expect that $\lambda_1$ for surface waves is of the fourth order in
the wave amplitude. Similarly, we expect no exact cancelation of the
non-vanishing Gaussian terms for $\lambda$, in the sixth order in
the wave amplitude, see also below. Then we have the following order
of magnitude estimates: $\lambda_1\propto \Omega
_{\boldsymbol{k}}\epsilon^4$ and $\lambda\propto \Omega
_{\boldsymbol{k}}\epsilon^6$. It follows that surface wave
turbulence is also weakly compressible in the sense of the ratio
$\lambda/\lambda_1\ll 1$, which is of the second order in the wave
amplitude. The Kaplan-Yorke dimension of the particles attractor on
the surface is close to the space dimension: $D_{KY}\approx
2-\lambda/\lambda_1\approx 2$. For dimensionless exponents ${\tilde
\lambda}_1\equiv \lambda_1/\Omega _{\boldsymbol{k}}$ and ${\tilde
\lambda}\equiv \lambda/\Omega _{\boldsymbol{k}}$ we find the order
of magnitude relation ${\tilde \lambda}\sim {\tilde
\lambda}_1^{3/2}$ holding at small wave amplitudes. The above
estimates are supported by numerical simulations performed by
\cite{Umeki} in a similar problem with {\it standing} surface waves.
It was shown there that both $\lambda\ll \lambda_1$ and $2-D_{KY}\ll
1$ hold. Moreover, the numerical values of the dimensionless
exponents ${\tilde \lambda}$, ${\tilde \lambda}_1$ found there, can
be easily seen to be in agreement with the relation ${\tilde
\lambda}\sim {\tilde \lambda}_1^{3/2}$. The box-counting dimension
of the particles attractor on the surface was found very close to
$D_{KY}$, which we expect to hold for wave turbulence as well. The
detailed calculations of the exact expressions for $\lambda_1$ and
$\lambda$ are subjects for future work.

We believe that our conclusion on the weak clustering in surface
waves with a small amplitude is an important step in identifying
possible reasons for the clusters of the floaters observed on liquid
surfaces ubiquitously. Our results suggest that other mechanisms
need to be explored, such as the wave breaking and Langmuir
circulation, see e.g. \cite{Thorpe}, the interplay of waves and
currents, see \cite{FFV}, and others.

We are indebted to G. Falkovich for constant help and useful
discussions. We are grateful to M. G. Stepanov for helpful
discussions. M. V. thanks S. Nazarenko for a useful remark. This
work was supported by the Israeli Science Foundation.

\bibliographystyle{jfm}
\bibliography{FVfloaters_arxiv}

\begin{thebibliography}{33}
\expandafter\ifx\csname natexlab\endcsname\relax\def\natexlab#1{#1}\fi

\bibitem[Balk {\em et~al.\/}(2004)Balk, Falkovich \& Stepanov]{BFS}
{\sc Balk, A.~M., Falkovich, G. \& Stepanov, M.~G.} 2004 Growth of density
  inhomogeneities in a flow of wave turbulence. {\em Phys. Rev. Lett.\/} {\bf
  92}, 244504.

\bibitem[Balkovsky {\em et~al.\/}(2001)Balkovsky, Falkovich \& Fouxon]{BFF}
{\sc Balkovsky, E., Falkovich, G. \& Fouxon, A.} 2001 Intermittent distribution
  of inertial particles in turbulent flows. {\em Phys. Rev. Lett.\/} {\bf 86},
  2790.

\bibitem[Bandi {\em et~al.\/}(2006)Bandi, Goldburg \& jr]{Bandi}
{\sc Bandi, M.~M., Goldburg, W.~I. \& jr, G. R.~Cressman} 2006 Measurement of
  entropy production rate in compressible turbulence. {\em Europhys.\ Lett.\/}
  {\bf 76}, 595--601.

\bibitem[Batchelor(2002)]{Batchelor}
{\sc Batchelor, G.~K.} 2002 {\em An Introduction to Fluid Dynamics\/}.
  Cambridge: Cambridge University.

\bibitem[Bec {\em et~al.\/}(2004)Bec, Gaw\c{e}dzki \& Horvai]{BGH}
{\sc Bec, J., Gaw\c{e}dzki, K. \& Horvai, P.} 2004 Multifractal clustering in
  compressible flows. {\em Phys. Rev. Lett.\/} {\bf 92}, 224501.

\bibitem[Boffetta {\em et~al.\/}(2006{\natexlab{{\em a\/}}})Boffetta, Davoudi,
  Eckhardt \& Schumacher]{BDES}
{\sc Boffetta, G., Davoudi, J., Eckhardt, B. \& Schumacher, J.}
  2006{\natexlab{{\em a\/}}} Lagrangian tracers on a surface flow: The role of
  time correlations. {\em Phus.\ Rev.\ Lett.\/} {\bf 93}, 134501.

\bibitem[Boffetta {\em et~al.\/}(2006{\natexlab{{\em b\/}}})Boffetta, Davoudi
  \& Lillo]{BDD}
{\sc Boffetta, G., Davoudi, J. \& Lillo, F.~De} 2006{\natexlab{{\em b\/}}}
  Multifractal clustering of passive tracers on a surface flow. {\em Europhys.
  Lett.\/} {\bf 74}, 62.

\bibitem[Bohr \& Hansen(1996)]{BH}
{\sc Bohr, T. \& Hansen, J.L.} 1996 Chaotic particle motion under linear
  surface waves. {\em Chaos\/} {\bf 6}, 554--563.

\bibitem[Cressman {\em et~al.\/}(2004)Cressman, Davoudi, Goldburg \&
  Schumacher]{CDGS}
{\sc Cressman, J.~R., Davoudi, J., Goldburg, W. \& Schumacher, J.} 2004
  Eulerian and lagrangian studies in surface flow turbulence. {\em New J.
  Physics\/} {\bf 6}, 53.

\bibitem[Cressman \& Goldburg(2003)]{CG}
{\sc Cressman, J.~R. \& Goldburg, W.} 2003 Compressible flow: Turbulence at the
  surface. {\em J. Stat. Phys.\/} {\bf 113}, 875--883.

\bibitem[Denissenko {\em et~al.\/}(2006)Denissenko, Falkovich \&
  Lukaschuk]{DFL}
{\sc Denissenko, P., Falkovich, G. \& Lukaschuk, S.} 2006 How waves affect the
  distribution of particles that float on a liquid surface. {\em Phys. Rev.
  Lett.\/} {\bf 97}, 244501.

\bibitem[Dorfman(1999)]{D}
{\sc Dorfman, J.} 1999 {\em Introduction to Chaos in Nonequilibrium Statistical
  Mechanics\/}. Cambridge Univ. Press.

\bibitem[Eckhardt \& Schumacher(2001)]{Bruno}
{\sc Eckhardt, B. \& Schumacher, J.} 2001 Turbulence and passive scalar
  transport in a free-slip surface. {\em Phys. Rev. E\/} {\bf 64}, 016314.

\bibitem[Falkovich \& Fouxon(2003)]{FF1}
{\sc Falkovich, G. \& Fouxon, A.} 2003 nlin.cd/0312033.

\bibitem[Falkovich \& Fouxon(2004)]{FF}
{\sc Falkovich, G. \& Fouxon, A.} 2004 Entropy production and extraction in
  dynamical systems and turbulence. {\em New J. Physics\/} {\bf 6}.

\bibitem[Falkovich {\em et~al.\/}(2001)Falkovich, Gaw\c{e}dzki \&
  Vergassola]{FGV}
{\sc Falkovich, G., Gaw\c{e}dzki, K. \& Vergassola, M.} 2001 Particles and
  fields in fluid turbulence. {\em Rev. Mod. Phys.\/} {\bf 73}, 913.

\bibitem[Nameson {\em et~al.\/}(1996)Nameson, Antonsen \& Ott]{NAO}
{\sc Nameson, A., Antonsen, T. \& Ott, E.} 1996 Power law wave number spectra
  of fractal particle distributions advected by flowing fluids. {\em Phys.
  Fluids\/} {\bf 8}, 2426--2434.

\bibitem[Ott(2002)]{O}
{\sc Ott, E.} 2002 {\em Chaos in Dynamical Systems\/}. Cambridge: Cambridge
  University Press.

\bibitem[Ramshankar {\em et~al.\/}(1990)Ramshankar, Berlin \& Gollub]{gorlum}
{\sc Ramshankar, R., Berlin, D. \& Gollub, J.} 1990 Transport by capillary
  waves. part i. particle trajectories. {\em Phys. Fluids~A\/} {\bf 2},
  1955--1965.

\bibitem[Ruelle(1996)]{R1}
{\sc Ruelle, D.} 1996 Positivity of entropy production in nonequilibrium
  statistical mechanics. {\em J.\ Stat.\ Phys.\/} {\bf 85}, 1--23.

\bibitem[Ruelle(1997)]{R2}
{\sc Ruelle, D.} 1997 Positivity of entropy production in the presence of a
  random thermostat. {\em J.\ Stat.\ Phys.\/} {\bf 86}, 935--990.

\bibitem[Ruelle(1999)]{R}
{\sc Ruelle, D.} 1999 Smooth dynamics and new theoretical ideas in
  nonequilibrium statistical mechanics. {\em J.\ Stat.\ Phys.\/} {\bf 95},
  393--468.

\bibitem[{Schroder~et~al}(1996)]{alstrom}
{\sc {Schroder~et~al}, E.} 1996 Relative particle motion in capillary waves.
  {\em Phys. Rev. Lett.\/} {\bf 76}, 4717.

\bibitem[Sommerer(1996)]{Sommerer}
{\sc Sommerer, J.~C.} 1996 Experimental evidence for power-law wave number
  spectra of fractal tracer distributions in a complicated surface flow. {\em
  Phys. Fluids\/} {\bf 8}, 2441--2446.

\bibitem[Sommerer \& Ott(1993)]{SO}
{\sc Sommerer, J.~C. \& Ott, E.} 1993 Particles floating on a moving fluid: A
  dynamically comprehensible physical fracta. {\em Science\/} {\bf 259},
  335--339.

\bibitem[Stepanov(2006)]{MGS}
{\sc Stepanov, M.~G.} 2006 private communication .

\bibitem[Thorpe(2005)]{Thorpe}
{\sc Thorpe, S.~A.} 2005 {\em The Turbulent Ocean\/}. Cambridge Univ. Press.

\bibitem[Umeki(1992)]{Umeki}
{\sc Umeki, M.} 1992 Lagrangian motion of fluid particles induced by
  three-dimensional standing surface waves. {\em Phys. Fluids A\/} {\bf 4},
  1968--1978.

\bibitem[Vucelja {\em et~al.\/}(2006)Vucelja, Falkovich \& Fouxon]{FFV}
{\sc Vucelja, M., Falkovich, G. \& Fouxon, I.} 2006 Clustering of matter in
  waves and currents. {\em submitted to Phys.\ Rev.\ Lett.\/} .

\bibitem[Yu {\em et~al.\/}(1991)Yu, Ott \& Chen]{Yu}
{\sc Yu, L., Ott, E. \& Chen, Q.} 1991 Fractal distribution of floaters on a
  fluid surface and the transition to chaos for random maps. {\em Physica D\/}
  {\bf 53}, 102--124.

\bibitem[Zakharov(1966)]{ZPhD}
{\sc Zakharov, V.~E.} 1966 PhD thesis, { \it in Russian}, Institute for Nuclear
  Physics (Physics and Mathematics), Novosibirsk.

\bibitem[Zakharov(1968)]{Z2}
{\sc Zakharov, V.~E.} 1968 Stability of periodic waves of finite amplitude on
  the surface of a deep fluid { (note a sign misprint in (1.8))}. {\em J. Appl.
  Mech. Tech. Phys.\/} {\bf 9}, 190--194.

\bibitem[Zakharov {\em et~al.\/}(1992)Zakharov, L'vov \& Falkovich]{ZLF}
{\sc Zakharov, V.~E., L'vov, V. \& Falkovich, G.} 1992 {\em Kolmogorov spectra
  of turbulence\/}. Berlin: Springer-Verlag.

\end{thebibliography}

\end{document}